\documentclass[a4paper,12pt]{article}
\usepackage{amsmath} 
\usepackage{graphicx}
\usepackage{amssymb}
\usepackage{setspace}
\usepackage{lineno}
\title{Frustration -- induced inherent instability and growth oscillations in pollen tubes}
\author{Mariusz Pietruszka\footnote{Corresponding author: \texttt{mariusz.pietruszka@us.edu.pl}} \\
{\footnotesize Laboratory of Plant Physiology, Faculty of Biology and Environmental Protection}\\
{\footnotesize University of Silesia, ul. Jagiello\'nska 28, PL-40032 Katowice, Poland}\\
{\footnotesize tel. +48322009453, E-mail: \texttt{mariusz.pietruszka@us.edu.pl}}
} 
\linenumbers
\begin{document}
\maketitle
\noindent
{\bf \abstractname{}.}
In a seed plant a pollen tube is a vessel that transports male gamete
cells to an ovule to achieve fertilization. It consists of one
elongated cell, which exhibits growth oscillations, until
it bursts completing its function. Up till now, the mechanism behind
the periodic character of the growth has not been fully understood. 
It is shown that the mechanism of pressure -- induced {\em symmetry frustration} occurring in the
wall at the perimeter of cylindrical and approximately hemispherical parts of a growing pollen cell, 
together with the addition of cell wall material,
suffices to release and sustain
mechanical self-oscillations and cell extension in pollen tubes. At the 
transition zone 
where symmetry frustration occurs and one cannot distinguish either of
the involved symmetries, a kind of 'entangled state' appears where
either single or both symmetry(ies) can be realized by the system. 
We anticipate that testifiable predictions made by the model ($f \propto \sqrt{P}$) may deliver, after calibration, a new tool to estimate 
turgor pressure $P$ from oscillation frequency $f$ of the periodically growing cell. Since the mechanical principles apply to all turgor regulated walled cells including those of plant, fungal and bacterial origin, the relevance of this work is not limited to the case of the pollen tube.
\section{Introduction}
\subsection{General outline}
The pollen tube has become a widely used cellular model system. In addition to being the fastest growing plant cell, it features periodic oscillations of the growth rate that have attracted numerous attempts to model the process\footnote{The origin of this oscillation is still unclear, though all hypotheses agree in that, the cell wall mechanics are essential to the oscillation (Chavarria-Krauser and Yejie, 2011).}. While recent models have increasingly incorporated biological features such as ion transport and intracellular trafficking, a simple feature with potentially significant impact has been overlooked in past approaches: geometry. We modeled the strain rates in the cell wall caused by turgor pressure depending on the different symmetries present in the pollen tube and found that a crucial area on the cellular surface of the pollen tube is characterized by so termed {\em symmetry frustration}. This area represents the transition zone between hemisphere-shaped apex and cylindrical shank. From a biological point of view this zone is crucial since numerous molecular landmarks of polar growth are present on one side of this zone and absent from the other. 

The model predicts that the transition zone undergoes local peaks in strain rate opening intriguing avenues for research focusing on the polarity of the growth process. Furthermore, we propose that changes between different symmetry regimes might be the mechanical underpinning of periodic changes in growth rate and shape observed during oscillatory growth. We believe that our model make an important contribution to the field of plant cytomechanics in general and pollen tube growth in particular.

\subsection{Preliminaries}
Pollen tubes are rapidly growing plant cells whose morphogenesis is largely determined by spatial gradients in the biochemical composition of the cell wall.
Pollen tube growth is a critical process in the life cycle of higher plants (Winship et al., 2010). It has garnered a lot of attention and is at the center of considerable controversy (Kroeger and Geitmann, 2011a). The pollen tube is the carrier of the male gametes in flowering plants. A controversy swirls around the modes of extension leading to periodicity in growth and growth rate (Winship et al., 2010). While some authors claim that hydrodynamics is the central integrator of pollen tube growth (Zonia and Munnik, 2007, 2009, 2011) leading to growth oscillations, the other couple the periodicity in growth dynamics to the changes in the wall material properties (Winship et al., 2010, 2011; Kroeger et al., 2011).
 
Pollen tubes are tip growing cells, which means that cell wall  unidirectional expansion is confined to the apex of the cell.  They display extremely rapid growth that can be also reproduced under {\emph {in vitro}} conditions. All growth activity - delivery of new cell wall material and cell wall deformation - occurs at the tip of the cell (Geitmann and Cresti, 1998): the tube is capped by an approximately hemispherically shaped dome, the apex, to which all growth activity is confined (Fig. 1A). The deformation is driven by the turgor pressure, a hydrodynamic pressure inside the cell. Interestingly, pollen tube evolution displays characteristic oscillations in growth and growth rate
(Plyushch et al., 1995; Hepler et al., 2001; Feijo et al., 2001).
The pollen tube (e.g. {\em Lilium longiflorum}, {\em Nicotiana tobacum}) growth oscillation depends on many underlying phenomena, amongst others the ion and mass fluxes, wall mechanical properties, system symmetry and turgor pressure. In isotonic conditions 
(Zonia and Munnik, 2007)
the average growth cycle period $T$ is about 50 s, while upon shifts to hypertonic or hypotonic conditions it is about 100 s and 25 s, respectively. The latter produce oscillations with typical frequencies ($f=1/T$): hypertonic -- 0.01 Hz, isotonic -- 0.02 Hz and hypotonic -- 0.04 Hz.
The longitudinal and transversal oscillation power spectrum of an individual {\em Nicotiana tabacum} pollen tube (Haduch and Pietruszka, 2012) is visualized in Fig. 2, and can not solely be described by a double exponent model (Pietruszka, 2012; e.g. Fig. 8A), suitable for normal cell evolution, but incomplete for periodical growth.

The cell wall is one of the structural key players regulating plant cell growth since plant cell expansion depends on an interplay between intracellular pressures and the controlled yielding of the wall (Geitmann, 2009). The cell wall may be treated as a polymer with the property of viscoelasticity (colloquially "elasticity"), generally having notably low Young's modulus $ \epsilon $ and high yield strain, which is a normalized measure of deformation representing the displacement between particles in the body relative to a reference length, compared with other materials.
Cell wall polymers are amorphous polymers existing above their 'glass' transition temperature, so that considerable segmental motion is possible. At ambient temperatures, the cell wall is thus relatively soft ($\epsilon \sim 1$ MPa) and easily deformed. Cells can grow to some extent simply by stretching their walls as they take up water. However, continued cell expansion involves synthesis of new wall material. Synthesis of cellulose at the plasma membrane and pectin and hemicelluloses components with Golgi apparatus deposits layers on the inside of the existing cell wall.    
A mechanical prerequisite for the unidirectional growth of pollen tubes for the (scalar) hydrostatic pressure is a softer cell wall at the tip of the cell, and more rigid at the distal part 
(Geitmann and Steer, 2006; Fayant et al., 2010). This gradient of mechanical properties is generated by the absence or scarcity of callose and cellulose at the tip 
(Aouar et al., 2009)
as well as by the relatively high degree of esterification of the pectin polymers in this region. The gradient in cell wall composition from apical esterified to distal de-esterified is reported to be correlated with an increase in the degree of cell wall rigidity and a decrease of visco-elasticity (Parre and Geitmann, 2005). Also microindentation studies show (Fig. 2 in  Winship et al., 2010) that the pollen tube tip is less rigid and that the distal stiffness may be opposed to apical softness.
Needless to say, to sustain growth processes, a balance between  the mechanical deformation of the viscoelastic cell wall and the addition of new cell wall material must be achieved 
(Kroeger and Geitmann, 2011b).
 
Turgor pressure is the pressure of the cell sap against the wall in plant cells. This is a force exerted outward on a plant cell wall by the water and solutes contained in the cell. As a result it gives the cell rigidity. An excess turgor pressure or cell wall local weakening can result in the bursting of a cell. Both constituents, turgor pressure and the wall properties are decisive for the mechanical properties and dynamics of the developing plant cell. The osmotically maintained (hydrodynamic) turgor pressure in living plant cells and the mechanical properties of the cell wall itself are among the most fundamental physical factors that dictate both cell growth and cell morphogenesis in plants (Schopfer, 2006). In fact, the physical properties of the wall and the turgor pressure have pivotal functions since they represent the 'downstream parameters' of all cellular signaling events 
(Chebli and Geitmann, 2007).
For our future purposes we note that turgor pressure is high in pollen tubes: 0.1 -- 0.4 MPa in lily (Benkert et al. 1997; Winship et al., 2010).

The pollen tube geometry can be described by two different symmetries - a hemisphere shaped apex, and a cylindrical shank zone, that are connected by a transition zone between the two parts. In the present paper we propose a mechanism of {\em symmetry frustration} occuring in this transition zone between the two involved symmetries as a possible mechanism responsible for growth rate oscillations. In simple terms by symmetry frustration we mean that a small ring of cell wall (hereafter referred to as interface $\Gamma$, Fig. 1B) is unable to 'decide' if it should behave as an elastic cylinder or an elastic sphere. Following this hypothesis, oscillations may arise because this mesoscopic ring behaves as if it 'jumps' periodically between the two mechanical states of different capacity of strain energy.

The application of growth tensors to developing plant organs has been known for a long time 
(Kutschera, 1989). Such mathematical description has been utilized to apical meristems where the proliferating cells produce tissue stresses, which in turn influence the structure of the developing organ and, hence the principal directions of growth 
(Hejnowicz, 1984). A different situation is observed for elongating plant organs, such as coleoptyles or hypocotyles, or elongation zones in roots, where cell division rarely takes place. Also, the directional evolution of a single cell is primarily observed for the elongating pollen tube. As it may be expected, various stresses occur in the elongating cell because of  different properties of cell walls exposed to the turgor pressure maintained by the gradient in the water potential
(Kutschera, 2000). The distribution of the wall stresses as well as deformation of the particular wall layers can be calculated by solving equilibrium equations of elasticity theory.
Such equations may help to locate deformation of a cell wall, exposed to an internal turgor pressure. The equilibrium equation may be derived both for materials deformed elastically (deformation vanishes when force equals zero) or non-elastically (plastic deformation survives, even when the acting force is removed). In this article we concentrate initially on elastic properties, because the highly controversial and uncertain mechanism of the oscillatory growth of pollen tubes is our main concern\footnote{Whereas the hydrodynamic model as it is used by Zonia et al. proposes gradual increase in turgor until a threshold when rupture of individual links between cell wall polymers occurs, Winship and coworkers (Winship et al., 2010) state that turgor is essentially stable, but an exocytosis--induced relaxation of the wall causes expansion. They postuale that variations in cell wall mechanical properties cause the oscillations and that variations in turgor (if there are any) are a passive consequence due to cell wall relaxation.} 
(Zonia and Munnik, 2011;  Kroeger et al., 2011; Winship et al., 2011). Though, plastic properties are inherently present in the proposed model {\em via} the derived (anharmonic) 'frustration potential', and the assumed cell wall building processes located in the sub-apical, annular region, presumably at (about) the $\Gamma$ -- interface  (Zonia and Munnik, 2008; Geitmann and Ortega, 2009).

Possible mechanisms have been proposed to account for the oscillation of pollen tube growth rate in quantitative terms 
(Bartnicki-Garcia et al., 2000; Feijo et al., 2001; Dumais et al., 2006). A model for calcium dependent oscillatory growth in pollen tubes has been put forward (Kroeger et al., 2008). More recently the finite elment technique was used (Fayant et al., 2010)
to establish biomechanical model of polar growth in walled cells. Also, the chemically mediated mechanism of mechanical expansion of the pollen tube cell wall by which deposition causes turnover of cell wall cross-links thereby facilitating mechanical deformation was set forth (Rojas et al., 2011).
The role of wall ageing in self-regulation in tip-growth was considered in (Eggen et al., 2011),
while a model of plasma membrane flow and cytosis regulation in growing pollen tubes was discussed in (Chavarria-Krauser and Yejie, 2011).  The irreversible expansion of the cell wall during growth as the extension of an inhomogeneous viscous fluid shell under the action of turgor pressure, fed by a material source in the neighborhood of the growing tip was examined in (Campas and Mahadevan, 2009). However, none of the models produced oscillations on mechanical basis\footnote{The model presented by Chavarria-Krauser and Yeije (2011), actually does not describe the cell wall, and hence, does not predict oscillations. The authors simply assume that the growth oscillations are given to understand the phase angle differences between growth velocity and the regulating mechanisms. However, by definition, oscillation is the repetitive variation, typically in time, of some measure {\em about} a central value (often a point of equilibrium), and pollen tubes do not exhibit such form of oscillations.
In fact, what we observe in pollen tubes is a periodical elongation (without shrinking phase), and can be a result of transitions between two or more different states, which is another definition of oscillation, which we adopt in this paper.}. 

In our approach, which does not contradicts previous achievements,  we explore the relationship between turgor pressure and nontrivial cell geometry by changing loss of stability picture 
(Wei and Linthilac, 2007)
to encompass cell wall mechanical properties in cylindrical and spherical geometries, both present in rapidly extending pollen tube. We base our physical model on the parametrized description of a tip growing cell that allows the manipulation of cell size, cell geometries, cell wall thickness, and local mechanical properties. However, the mechanical load (contrary to op. cit.) is applied in the form of hydrostatic (constant) pressure. 

An important feature of pollen tubes elongation is that growth rate oscillates and, additionally, many of the underlying processes also oscillate with the same period, but usually with different phase (e.g. Fig. 1c in (Zonia and Munnik, 2011)). However, the role of the oscillating ion gradients and fluxes in the control of pollen tube growth (Hepler and Winship, 2010) 
is beyond the scope of this paper and we will not discuss it here. Nonetheless, the outlined scenario leaves space for the periodical ion and mass fluxes in and out of the cell. We also share the fundamental view expressed in (Rojas et al., 2011;  
Proseus and Boyer, 2006)
 that the wall extension is primarily a biophysical (mechanical) process, although ultimately dependent on enzymatic activity, and that under conditions where the enzymatic background can be subtracted the biophysical process still proceedes normally.
 
Any new model should deliver testifiable and quantitative predictions that can be validated by experimental data. In case of pollen tubes, it is necessary to present predictions that go beyond stress values which are inherently difficult to measure. The presented model satisfies this requirement offering, among others, an experimentally testifiable power law ($\omega \propto \sqrt{P}$) between the turgor pressure $P$ and growth oscillation  angular frequency $\omega$.
\section{Results and discussion}
In search for the cause of experimentally observed pollen tube growth oscillations we link analytic stress/strain relations with the mechanical properties of a tip growing cell, located at the perimeter of hemispherical dome and cylindrical shank. This is based on the observation that cell wall assembly by exocytosis occurs mainly at an annular region around the pole of the cell 
(Geitmann and Dumais, 2009; Zonia and Munnik, 2009)
and that the concomittant turgor driven deformation of the cell wall causes characteristic strain in the hemisphere shaped apex of the cell (Fayant et al, 2010; Rojas et al., 2011). 
The dynamical properties of such a complex growing system should be found self-consistently (meaning that the turgor pressure and the wall mechanical properties are conjugate magnitudes that usually form coupled equations, which have to be solved by iteration methods). Nevertheless, in first approximation the following heuristic solution can be proposed.

Assuming an intrinsic turgor pressure $P$, and a much smaller external pressure, of yet unspecified origin $ p_{\mathrm{ext}} $ (it can be just atmospheric pressure) producing an effective pressure $ \tilde{P} = P+ p_{\mathrm{ext}}$) the 
equilibrium equation for the displacement vector $ \vec{u} $, which is the shortest distance from the initial to the final position of a moving point, takes the form (Landau and Lifshitz, 1986):
\begin{equation}
2\;(1-\nu)\;
\mathrm{grad}(\mathrm{div} \;\vec{u})
-(1-2\nu)\; \mathrm{curl}(\mathrm{curl}\;\vec{u})=\vec{0}
\end{equation}
where $ \nu $ is the Poisson coefficient, which is the ratio, when a sample object is stretched, of the contraction or transverse strain, to the extension or axial strain. From now on Young modulus $\epsilon(z)$ and Poisson coeficient $\nu$ are assumed as picewise constant functions, so they remain constant on the interface $\Gamma$.

Note, that by acting divergence operator on both sides of Eq. (1) we receive $\bigtriangleup\; \mathrm{div}\; \vec{u}=0$, i.e. $ \mathrm{div}\;\vec{u} $ denoting the volume change due to displacement field is a harmonic function satisfying Laplace's equation.

Eq. (1) may be solved analytically, providing that a problem exhibits a high degree of symmetry. In particular it can be solved exactly for spherical and cylindrical symmetries. Both symmetries are present in the description of pollen tube self-similar elongation, since the distal part resembles cylindrical tube while the apex is a hemispherical dome (Fig. 1B). 
Thus, the symmetries present in both subdomains should be utilized in the description of pollen tube shape and dynamical properties.

By assuming cylindrical symmetry (for the displacement vector field $ \vec{u}=(u_r,u_\phi,u_z)=(u_r,0,0) $, which is obtained under the assumption that the total length of the cylinder part remains constant (the axial elongation of the more rigid shank upon application of a constant internal pressure $P$ we assume as negligible), and hence we accept $u_z=0$ in the Ansatz $(u_r,u_\phi,u_z)=(u_r,0,0)$ instead of $(u_r,u_\phi,u_z)=(u_r,0,u_z)$, which would lead to unavoidable numerical solution of Eq. (1)), and representing field operators (grad, div and curl)  in cylindrical (polar) coordinates, Eq. (1) can be reduced to a much simpler form:
\begin{equation}
\frac{d}{dr}\left[\frac{1}{r}\frac{d}{dr}(ru_r) \right]=0.
\end{equation}
This differential equation  can be solved for the displacement field $ u_r $ to yield the displacement for the cylindrical symmetry
\begin{equation}
u_r^c=ar+\frac{b}{r},
\end{equation}
where $ a $ and $ b $ are constants to be determined from the boundary conditions (Landau and Lifshitz, 1986). 
Also, by introducing spherical coordinates with the origin in the center of a sphere the displacement field $ \vec{u} $ is a function of the radius $ r $: $ \vec{u}=(u_r,u_\theta,u_\phi)=(u_r,0,0) $. Therefore $ \mathrm{rot}\; \vec{u}=0  $ and Eq. (1) reads: $ \mathrm{grad}\; \mathrm{div} \; \vec{u}=0 $. Hence, for the spherical symmetry we receive for displacement
\begin{equation}
u_r^s=ar+\frac{b}{r^2},
\end{equation}
where the upper index in Eqs (3) and (4) has been substituted to differentiate solutions for cylindrical (c) and spherical (s) geometries.
The geometry of the elongating pollen tube can be described with a cylinder of radius $ r_T $ capped by a half prolate spheroid with short radius $ r_T $ and a long radius $ r_L $ (see e.g. Fig. 1a in (Fayant et al., 2010)).
Thus, as a model for normally growing tube we propose a thin-walled hollow cylinder ended by a hemispherical shell immersed in an external pool of pressure $ p_\mathrm{ext} $ and filled with a cell sap with turgor pressure $ P $ (we equate both radii $ r_T=r_L $, for simplicity). The inner radius of the cylinder and a sphere is $ r_1 $, while the outer radius is $ r_2 $ (Fig. 1B). 
Another simplifying assumption of the model is that we deal with weak (elastic) interactions at the interface $\Gamma$, due to wall building processes occuring at this region, and hence the deflection field may slightly differ on both sides (from the mechanical point of view they may be treated as weakly coupled). It is consistent with the view that deposition chemically 
loosens the wall by breaking load--bearing cross--links 
while simultaneously creating new, load--free cross--links, 
thereby effecting a fail-safe scenario for mechanical 
expansion (Rojas et al., 2011).                                                                                                                                                                                                                                                                                                                                                                                                                                                                                                                                                                

Based on the proposition that the mechanical cell wall properties at the growing tip must be different from those in the shank, it was suggested (Dumais et al., 2004; Kroeger et al., 2008) that an anisotropy in the cell wall elasticity is required to account for the transition between spherical and tubular shape at the tip of the cell. Also, it was found (Geitmann and Parre, 2004) that the rigidity of the tip of the pollen cell was an increasing function of the distance from the apex. 
Therefore, the elastic properties of the cylinder imitating the cell wall at the shank and the hemispherical tip are represented by two pairs of material constants: Young's modulus $ \epsilon $, also known as the tensile modulus, which is a measure of the stiffness of an elastic material and is a quantity used to characterize materials, and Poisson coefficient $ \nu $. In further calculations, we assume different values for these coefficients for distal (thick and rigid) and apical (thin and elastic) walls of a pollen tube. Such assumptions, about different values of mechanical constants
at the apical dome and cylindrical shank, are fully justified (see e.g. Fig. 4 in (Geitmann and Parre, 2004) , where a spatial distribution of the Young's modulus is presented). Because we consider relatively small elastic deformations, the stress and strain tensors are related by the Hooke's law of elasticity (which is an approximation that states that the extension of a spring is in direct proportion with the load applied to it) and makes the deformation reversible.  For the radial part of the stress tensor $ \sigma_{ij}  $ we have: $ \sigma_{rr}=-p_{\mathrm{ext}}  $ at $ r=r_2 $ and $ \sigma_{rr}=-P  $ at $ r=r_1 $. Since the off-diagonal elements vanish, we are left with the strain
$ u_{rr}=\frac{du_r}{dr}=a-\frac{b}{r^2} $, $ u_{\phi \phi}=\frac{u_r}{r}=a+\frac{b}{r^2} $ and $ u_{zz}=0 $, the interesting radial $ \sigma_{rr} $ element of the stress tensor reads\footnote{In fact, to receive smooth solutions on $\Gamma$ -- interface equations for different geometries  should be connected by transmission (gluing) conditions equating the forces and deflections on each side: $\sigma \mathbf{n}=\sigma' \mathbf{n'}$ and $\mathbf{u}=\mathbf{u'}$, where $\mathbf{n}$ denotes the exterior normal to the boundary. 
In first approximation we let both subdomains be weakly coupled (visco -- plastic phase) while cyclic wall building processes take place at $\Gamma$, and strongly coupled mechanically (visco -- elastic phase) when wall building processes expire. We consider only the radial part of the stress $\sigma_{rr}$, deflection $u_r$ and strain $u_{rr}=\partial_r u_r$ tensors on $\Gamma$, for simplicity. This Ansatz, however, does not qualitatively influence the results.}:
\begin{equation}
\sigma_{rr}=\frac{\epsilon}{(1+\nu)(1-2 \nu)}
\left[a-(1-2 \nu)\frac{b}{r^2} \right].
\end{equation}
By assuming boundary conditions as above, parameters  $ a $ and $ b $ can be calculated (Lewicka and Pietruszka, 2009). They both depend on material constants: Young's modulus $ \epsilon $ and Poisson coefficient $ \nu $, cylinder geometry ($ r_{1}$, $r_{2}$ -- radii) and pressure values $ P $ and $ p_{ext}$. 

Quantitative calculations steming from Eqs (3) -- (5) are presented in Figs 3 -- 4.
In Fig. 3A we observe (a) lowering of  deformation $u_r$ with radius $r$ of apical hemispherical shell  (b) almost constant $u_r$ for a thin-walled cylinder. This fact produces a wall stress presented in Fig. 3B equal to about 60 MPa (for the model parameters), which seems big enough to cause cell wall instability at $\Gamma$ -- interface\footnote{Typical plant cell turgor pressures in the range of 0.3 to 1.0 MPa translate into between 10 and 100 MPa in the walls (Wei and Linthilac, 2007).}. The calculated wall stress of the order of tenths of MPa is enough to cause local wall instability (radial strain), subsequent axial relaxation, wall building and unilateral cell expansion at the interface between the hemispherical apex and cylindrical shank. Unavoidable repetition of this process, owing to a constant effective pressure and a positive feedback mechanism necessary to drive oscillations (to overcome damping due to viscosity), may generate observable oscillations which continue until the wall building processes expire. 
Depending on the wall thickness, the calculated stress is equal to the distance between the curves (a) and (b) (inset). 
The tensile stress difference calculated at the apex  and the distal part of the pollen tube cell wall ($\sigma_{rr}(\mathrm{apex})-\sigma_{rr}(\mathrm{distal})$) is shown in Fig. 3C. Here it is parametrized by the turgor pressure $P$ acting on the cell wall. The tensile stress in the wall clearly rises, as we increase the turgor pressure -- this would, in turn, cause an increased growth oscillation frequency, as it is observed experimentally  in the transition from  hypertonic, through isotonic to hypotonic conditions (Zonia and Munnik, 2007). 

The influence of turgor on the oscillation period, as predicted by the model described in (Kroeger and Geitmann, 2011a), is also in agreement with our results presented in Fig. 4A and their consequences: the higher the pressure, the higher the oscillation frequency. Tensile stress difference at the transition zone, parametrized by the wall thickness, is presented in Fig. 4B. 

From the expression for tensile stress difference due to the symmetry change (note, curvature discontinuity and stress singularity occur at the transition zone) $\sigma_{rr}^c-\sigma_{rr}^s \equiv \sigma_{rr}(\mathrm{cylinder})-\sigma_{rr}(\mathrm{sphere})$ which equals
\begin{equation}
\sigma_{rr}^c-\sigma_{rr}^s=P
\left[
\frac{ {r_1}^3 {r_2}^3}{r^3
   \left({r_2}^3-{r_1}^3\right)}-\frac{ {r_1}^2
   {r_2}^2}{r^2
   \left({r_2}^2-{r_1}^2\right)}-\frac{
   {r_1}^3}{{r_2}^3-{r_1}^3}+\frac{
   {r_1}^2}{{r_2}^2-{r_1}^2}
   \right]
\end{equation}
and from the opposite formula: $\sigma_{rr}^s-\sigma_{rr}^c$, it is shown that calculations performed for an infinitesimally narrow ring $\Gamma$, where both geometries (cylindrical and spherical) intervene, a {\em symmetry frustration} -- leading to oscillations of the radial part of the stress tensor -- may take place. 'Frustration' originates from the fact that none of the locally involved symmetries is distinguished. 
On the other hand, the calculated from Eq. (6) (by evaluating $E=\int\sigma_{rr}dr$) 
strain energy density reads:
\begin{equation}
E_{\pm} \propto \pm P\;\frac{ {r_1}^2 {r_2}^2 \left[2 r^3+2 r
   \left({r_1}^2+{r_1}
   {r_2}+{r_2}^2\right)-{r_1} {r_2}
   ({r_1}+{r_2})\right]}{2 r^2
   \left({r_1}^4+{r_1}^3 {r_2}-{r_1}
   {r_2}^3-{r_2}^4\right)}
\end{equation} 
The quasi--discrete energy levels $E_-$ and $E_+$ (possesing, however, a small dispersion $\delta E\cong 0.0003$ [energy u.]) 
presented in Fig. 5  are non--degenerate due to the existence of a constant turgor pressure $P$ which leads to the observed splitting (see also upper inset). Still, since both levels originate from the symmetry change at the critical limit considered in this work, they can be attributed to the oscillations in the pollen tube growth functions. Thus, the {\em resonating frequency} of growth (growth rate) corresponds to the energy difference $E_{+}-E_{-}\cong 2E$ (since $E_{+}\cong -E_{-}$), which in turn is directly proportional to the turgor pressure $P$. Consequently, we may equate the transition energy $2E$ between the resonating levels (Fig. 5) with pollen tube oscillation frequency observed in experiments (Eq. (7) implies that if $P=0$ then the system exhibits no oscillations, which is exactly the case, see also the plot of the potential energy $U(r)$ at $r=r_0$ in Fig. 6).

Notwithstanding, we note that the considered effect is exclusively connected with geometrically induced stress in the wall which may be linked with symmetry frustration\footnote{A spatially degenerate ground state will undergo a geometrical distortion (the alteration of the original shape) that removes that degeneracy, because the distortion lowers the overall energy of the whole complex.}, and one can express it in measurable units [Pa  m]. Indeed, calculating definite integral over the function expressed by Eq. (6): $\int_{r_1}^{r_2} [...]dr $ (with $r_1=5$ and $r_2=5.25$ $\mu\mathrm{m}$) we receive the strain energy density: $E_{\pm}=\pm 0.256$ [MPa $\mu$m]. Therefore, the difference of strain energy density between the two levels is about 0.5 MPa for micrometer length scales typical for the width (which is about 250 nano-meters) of the pollen tube cell wall. 
Such energy density may lead to oscillations which are observable not only in growth rates. From our model it can also be deduced that the apical geometry oscillates (due to deformation $u_r$ located initially at $\Gamma$, compare Figs 3B and 4B in Pietruszka et al., 2012) to produce the so called pearled morphology (Rojas et al., 2011). Relicts (residues) of such deformations at $\Gamma$ are traced in Fig. 1c and 6a, c (ibid.) as crests smeared out on a distance $\lambda$, in agreement with our model\footnote{Such geometrical oscillations of the wave--length $\lambda$ will be obtained when frustration occurs, and the cylindrical and spherical symmetries will be present on $\Gamma$ -- contour interchangeably; compare Movie S1 and S2 in (Pietruszka et al., 2012).}. However, what else draws our attention is that there is no sign of deflecions at a distance shorter than the tube radius $R$. The latter observation further supports the main idea presented in this paper of specific role of the $\Gamma$ -- interface in initiating oscillations. By assuming, after Rojas et al. (2011),
the value of the linear $v_{\mathrm{avg}}=0.2 $ [$\mu$m/s] of the elongating cell and taking the average oscillation period $T=50$ s from Fig. 7  we receive the observed wavelength of about $\lambda=10$ [$\mu$m], which is a doubled value of the radius $R$, as it should be expected assuming the correctness of our approach. In this picture, the distance $\lambda$ comprise the local deflection/wall stress/stress relaxation/recovery through wall building processes for every period $T$.

Transitions between the states of different symmetry are shown in Fig. 5. The system is pumped with energy to jump over the analytic discontinuity (energy gap) between the (hemi-)spherical and cylindrical geometry. As a positive feedback mechanism necessary to drive the oscillations (to prevent damping due to viscosity) the energy is absorbed (ATP--pumping (Rounds et al., 2011)) in the transition zone above the state $E_{-}$ producing the exited state $E_{+}$. Then the system returns (by spontaneous symmetry breaking, to reduce the energy of the overall system) to the lower symmetry (cylindrical) state $E_{-}$ stimulating axial expansion.
Both transitions (up and down) close one growth cycle with the prediced transition's rate $\omega\propto\frac{2\pi}{T}$, where $T$ is the period.
The whole process is repeated at the expense of pressure $P$ and ATP--energy needed for wall synthesis (exhibiting growth oscillations), and eventually expires or
reaches critical instability (the cell bursts). 

%

In a mechanical anharmonic oscillator, the relationship between force and displacement is not linear but depends upon the amplitude of the displacement. 
The nonlinearity arises from the fact that the spring (here: cell wall) is not capable of exerting a restoring force that is proportional to its displacement because of stretching in the material comprising the wall. 
As a result of the nonlinearity, the vibration frequency and amplitude can change, depending upon the system elements displacement upon pressure $P$. 
An approximate derivation performed in Appendix 1 delivers the analytic form of the (dual) 'frustration potential', which is a sum of attractive and repulsive forces, possibly responsible for experimentally observed growth rate oscillations in pollen tubes. The latter, which is given by Eq. (12) and visualized in Fig. 6, we describe here shortly:
Pollen tube oscillations are trapped at the potential well about the equilibrium point $r_0$ for the corresponding symmetric (harmonic) potential. Oscillation frequencies and amplitudes of the anharmonic potential depend upon the turgor pressure values, see Eqs (6) and (7), as it is observed (e.g. Fig. 4 in Kroeger et al., 2011;  Kroeger and Geitmann, 2011a)). Wall expansion is allowed by molecular separation ($r$ -- values) exceeding those of harmonic potential.
Dissociation energy at zero potential level corresponds to system instability (burst at $\Gamma$). The constant turgor pressure, as taken from Fig. 5, induces the value of $U(r)$, hence the frequency of the oscillation and its amplitude. The low-lying (trapped) values deliver high frequencies and small amplitudes, while the higher-lying potential values -- low frequencies and larger amplitudes of oscillations. Above the critical threshold (corresponding to 'zero energy' at the vertical scale) a bond breaking occurs and the pollen tube burst at the transition zone, or deliver male gametes completing its function. The lower plot represents only one branch (of the prevailing cylindrical symmetry) of the full frustration potential; the second branch (above) is in 'dual' subspace, and the oscillations take place between both branches. For the negative values of the cylindrical symmetry, as shown in the plot, the mechanism of symmetry breaking favorizes this 'lower order' (cylindrical) symmetry for cell extension.

The presented in Fig. 6 frustration potential is a more convenient model for vibrational structure of wall constituing molecules than harmonic oscillator potential, because it explicitly includes the effects of bond breaking and accounts for anharmonicity of real bonds in the extending cell wall. It is also responsible for the inherent instability at the $\Gamma$ -- interface of a growing tube (and -- in consequnce -- polymer building process), which can be experimentally supported by the fact that the pollen tubes always rupture at the transition zone where the radial part of the strain tensor is  considerable (Pietruszka et al., 2012, Movie S3). The form of the potential also contributes to the long debate among plant physiologists about the elastic/inelastic extension of plant cell wall in simple terms:  any departure from parabola centered around $r_0$  will lead to plastic extension, corresponding to elongation growth (Fig. 6). In addition, the infinite potential barrier at low distance $r$ values prevents the growing cell wall from shrinking at a given pressure $P$.

In order to calculate  the  value of the resonance angular frequency $\omega$ we momentarily accept the approximate (classical) relation: $E \propto \omega^2$. Assuming $P=0.3$ MPa, taking the approximate $A$ constant from the fit (see Fig. 7) we receive $f\cong 0.09$ Hz, a value which belongs to the observed frequency spectrum in pollen tube growth functions ($0.01$ Hz -- $\sim 0.20$ Hz, (Haduch and Pietruszka, 2012) for tobacco, Fig. 2; and
(McKenna et al., 2009) for lily). 
As an aside, we stress that the calculated from Eq. (7) resonance frequency satisfies a power law $\omega \propto \sqrt{P}$ (see Appendix 2 for detailed derivation). The application of this important relation to the experimental data
is presented in Fig. 7.
It is easy to notice, that this relation ($\omega = 2\pi A \sqrt{P}$, or equivalently $f=A \sqrt{P}$, where $A$ is a constant -- connected with the wall mechanical properties -- to be determined from experiment, and [P] = MPa), if inverted, can serve (after calibration) to estimate difficult to measure turgor pressure $P$ values from easy to measure oscillation frequencies (or periods $T$).

Furthermore, it is clear that the material properties of the cell wall in the apical region should not be homogeneous, and therefore a proper mechanical description of growth must involve a gradient in material properties from the apical to the distal region (Fayant et al., 2010; Eggen et al., 2011). It has been shown (Eggen et al., 2011) that the calculated "expansion propensity" as a function of the distance from the apex measured in units of the tube radius $R$ (notation, as in (Eggen et al., 2011)) shrinks to an area near the apex. Closer examination reveals that the inflection point is located at about 1 pollen tube cylinder radius $R$, a place where we perform our calculations. The latter statement means that the slope is the greatest at $z \sim R$  (in axial direction). This, and the fact that the 'dilution' sector is shown (Fig. 2 in Eggen et al., 2011)) exactly at the limit of the two considered axisymmetric zones,  is consistent with the view of intense changes of the wall mechanical properties at the limit of the distal and apical part. The 'dilution' effect is caused in our model by a rapid surface expansion due to displacement $u_r$ of the $\Gamma$ -- interface (see also (Parre and Geitmann, 2005): Fig. 1 -- a local dip in the wall stiffness appears at about 10 $\mu$m from the apex, a place where our calculations are performed; and Fig. 7 (2) showing the possible localization of the transition zone from the cylindrical to spherical symmetry). Corresponding radial strain may trigger exocytosis that results in delivery of new cell wall material which rejuvenates this area. However, we should note, what we observe is not only according to the  mechanical properties gradient, but mainly due to the changing  symmetry at this place and the analytical consequences (curvature discontinuity) of this fact (compare e.g. Eqs 3 and 4).
Likewise, observations, together with the analysis of Flourescence Recovery After Photobleaching (FRAP) experiments  (Geitmann and Dumais, 2009; Bove et al., 2008) indicate that exocytosis is likely to occur predominantly in the same annular region (cf. (Geitmann, 2010), Fig. 1) where wall expansion rates are greatest. It is concluded in the same work, that tip growth in plant cells does not seem to happen exactly at the tip.  Further supporting data is provided  in (Zonia and Munnik, 2008), where the vesicle fusion with the plasma membrane was shown to approach to within 2-5 $\mu$m distal to the apex. 
The observations that growth is mainly at an annular region around the pole are in accord with calculations presented in this article.
Probing mechanical properties at the perimeter of cylindrical and spherical part may result in calculation of the local rate of exocytosis (the conventional explanation for this phenomenon, i.e. oscillating exocytosis rate (see e.g. Fig. 4A in (Cardenas et al., 2008),  Fig. 2 in (McKenna et al., 2009), is widely accepted). This may be roughly estimated by taking the oscillation frequency from Fig. 7.
The read off value: $f \sim 0.02\: -0.03$ Hz is in accord with the main observed periodical mode (Zonia and Munnik, 2007) in the longitudinal power spectrum of pollen tube oscillatory motion, and presumably may be equated with the rate of exocytosis and new cell wall assembly in {\em Nicotiana tobaccum} pollen tubes and temporal variations in the secretion of cell wall precursors (Kroeger and Geitmann, 2011a). 
This periodical growth activity, in turn, could be related (among others) to the relaxation of the stress (invoked by the hydrostatic turgor pressure) in the $\Gamma$ -- interface, i.e. in close proximity to the advancing apex of the cell, probably at the vesicle delivery zone in the subapex, where the vesicles are released into the cytoplasm  (Fig. 3 in (Kroeger et al., 2009)). In addition, even small stress/strain fluctuations at this narrow cylindrical ring, belonging to both adjacent zones, localized on its circumference could lead to macrosopically observable change of orientation of this ring and consequently direction change of the elongating pollen tube. This is indeed the case --  see e.g. Fig. 1F in  (Zonia and Munnik, 2008), and even more pronounced in Fig. 1 in  (Calder et al., 1997). It seems that polar growth in pollen tubes is associated with spatially confined dynamic changes in cell wall mechanical properties (Zerzour et al., 2009; and Appendix 2). 
Furthermore, the time course of an experiment, showing very little change in turgor pressure during cell growth, was measured by pressure probe monitoring of growing {\em Lilium longiflorum}  (Winship et al., 2010, Fig. 1a) and re-analysed in (Zonia and Munnik, 2011, Fig. 3b-c). Even though direct measurements fail to indicate large-scale turgor changes during growth, rapid small-scale pressure changes (jumps) are visible, which presumably\footnote{There is, however, no evidence for cellulose microfibrils to be involved in small changes in turgor. These small jumps in turgor may be simply imprecisions in the measurement method. This method requires readjusting the meniscus in a pressure probe needle for each data point and the measurement is inherently associated with significant noise.} may be caused by the change of orientation of the tilt angles of wall building cellulose microfibrils in subsequent growth cycles in the considered transition zone\footnote{The direction of maximal expansion rate is usually regulated by the direction of net alignment among cellulose microfibrils, which overcomes the prevailing stress anisotropy (Baskin, 2005).}. As stated (ibid.), the measured periodicity for pressure oscillations ranges from 12 s to 25 s, which is the same as the routinely reported for oscillatory dynamics in lily pollen tubes (McKenna et al., 2009; Zerzour et al., 2009) and approximately agrees with the calculated frequency ($f=1/12 \cong 0.09$ Hz) from our model (see next paragraph). Closer examination of Figs 3a-c in (Zonia and Munnik, 2011) reveals, in addition, a slight but steady diminishing of turgor (negative slope) which can be a consequence of cell volume expansion in every cycle. The latter observation may be associated with the passive role of the turgor pressure in pollen tube growth, at least at unchanging osmotic potentials.
In conlusion, we agree with the view, that in the growth process the (main) energy supply is derived from turgor, while the growth rate and direction from the wall local properties (Winship et al., 2010).

Expansive growth in a plant cell relies on the interplay between the internal turgor and the forces in the cell wall opposing deformation. Which of the two parameters controls the dynamics of growth has been controversial in the case of pollen tube growth (Zerzour et al., 2009). 
The long-standing model of pollen tube growth considers that cyclic changes in cell wall properties initiate growth (Holdaway-Clarke and Hepler, 2003; Winship et al., 2010). 
A new model based on accumulating data from recent work indicates 
that oscillations in hydrodynamic flow and intracellular pressure initiate growth (Zonia 2010; Zonia and Munnik, 2011). Both models agree that once growth initiated, osmotic pressure drives cell elongation. 
We have shown that the rapid polar growth phases during oscillatory growth in pollen tubes may be preceded by a strain --  induced softening of the cell wall at the brink of the apical and distal parts ($\Gamma$). We also showed that cellular turgor pressure does not need to undergo changes during these repeated growth phases to display periodicity in growth. However, turgor pressure still preserves an important role, together with the cell wall mechanical properties, in controlling the dynamics of pollen tube growth by changing wall stress and hence oscillation frequencies in the different osmotic environments -- the frequency of oscillatory pollen tube growth in our model can be altered by changing the osmotic potential value of the surrounding medium (Kroeger and Geitmann, 2011a).  

There are many observations of oscillations that could affect growth rates (such as wall material deposition and extracellular ion fluxes).
However, even from purely mechanical calculations, performed at the boundary between the wall  cylinder shell and hemispherical shell at the apex, the following picture for the sequence of events for the elongating pollen tube  emerges.
A given (constant) turgor pressure produces {\em different} strain at the apical and distal wall parts possessing various mechanical properties and different symmetries. This phenomenon is especially important at the narrow interface between these neighbouring parts. Consequently, a localized elevated  stress of the order of tenths MPa is invoked in the wall causing serious instability at the brink of both sections generated by symmetry frustration, and the cell wall relaxation (loosening) process takes place in order to reduce tensile stress. This initiates a wall building process (which is an implicit assumtion of the model) in meridional direction. However, the effective turgor pressure produces strain at the similar location (in the co-moving -- with the moving tip -- reference frame) axially equidistant from the tip and the whole process/cycle repeats. This eventually results in time -- periodicity in the growth dynamics recognized in the literature as pollen tube oscillations. 
\section{Final comment}
This article offers a nontrivial solution for a long -- sought mechanism of pollen tubes growth oscillations, which is the subject of swirling controversy in the field. It is based on the phenomenon of {\em symmetry frustration} of the cell wall in the apical region. This simple physical mechanism results in anlytically determined asymmetric 'frustration potential', the appearance of the landscape of 'discrete' energy levels at (different) constant pressures and the aforementioned oscillatory growth comes from the transitions between them. Moreover, a scaling relation between the turgor pressure $P$ and the angular frequency of the oscillations $\omega$ is derived, which is represented by a power 1/2 -- law ($\omega\propto \sqrt{P}$). Later on, this prediction is successfully verified against a real plant physiological experimental data. 
\subsection*{Acknowledgements}
Author thank dr Pawe\l{} Gusin, string theory specialist, Institute of Theoretical Physics, University of Wroc\l{}aw, Poland and Professor dr hab. Jan S\l{}adkowski, Department of Astrophysics and Cosmology, Institute of Physics, University of Silesia, Katowice, Poland for critical reading of the manuscript. I feel specially indebted to David Logan, 
Coulson Professor of Theoretical Chemistry
Oxford University, Physical and Theoretical Chemistry, Oxford, United Kingdom for expressing his kind opinion and encouragement about this work.
\subsection*{Appendix 1}
(i) The local force equation of motion in the mechanics of continuous bodies (due to Cauchy) reads (Lubliner, 2006):
\begin{equation}
\partial_j\sigma_{ij}+F_i=0,
\end{equation}
where i, j = 1, 2, 3, and Einstein's summation convention is used. 
By ignoring any possible shear stresses and assuming for the infinitesimally small $\Gamma$ -- interface $\sigma_{\phi \phi}=0$ and 
$\sigma_{zz}=0$
\begin{equation}
\sigma_{ij}=\left(
\begin{array}{ccc}
\sigma_{rr} & 0 & 0\\ 
0 & \sigma_{\phi\phi} & 0\\
0 & 0 & \sigma_{zz}
\end{array} 
\right)
\simeq\left(
\begin{array}{ccc}
\sigma_{rr} & 0 & 0\\ 
0 & 0 & 0\\
0 & 0 & 0
\end{array} 
\right)
\end{equation}
we receive $\sigma_{rr}=\sigma(r)$, which is the only matrix element that survives. From Eq. (8) we have
\begin{equation}
\int \partial_j \sigma_{ij}dx^i+\int F_i dx^i=0
\end{equation}
so since $\int F_i dx^i=-U(x)$ is the potential energy then $\int \partial_j \sigma_{ij}dx^i=U(x)$. Hence in our case $\int \partial_r \sigma_{rr}dr= \sigma_{rr}=U(r)$, and one gets: $U(r)=\sigma_{rr}=\sigma(r)$. 

\noindent
(ii) By identyfying 
\begin{equation}
U(r)=\sigma(r)
\end{equation}
we may follow Eq. (6) to receive
\begin{equation}
\Delta \sigma^{\pm}=\pm\frac{\alpha}{r^3}\mp\frac{\beta}{r^2}+C
\end{equation}
where
$\alpha=P\frac{(r_1 r_2)^3}{r_2^3-r_1^3}$, $\beta=P\frac{(r_1 r_2)^2}{r_2^2-r_1^2}$ and $C$ is a constant. Next, in order to find $r_0$ we write $\Delta \sigma'(r_0)=0$
to get
\begin{equation}
-\frac{3\alpha}{r^4}+\frac{2\beta}{r^3}=0
\end{equation}
Hence 
\begin{equation}
r_0=\frac{3}{2} \frac{\alpha}{\beta}.
\end{equation}
Since $U \equiv \Delta \sigma$ we can plot the 'symmetry frustrated' potential $U(r)$, Fig. 6.

\noindent
(iii) Considering small oscillations around equilibrium $r_0$, we may introduce new coordinate $\rho$: $r=r_0-\rho$ and by substituting it to Eq. (12) receive
\begin{equation}
U(x)=\frac{1}{r_0^3}\frac{\alpha}{(1-x)^3}-\frac{1}{r_0^2}\frac{\beta}{(1-x)^2}
\end{equation}
where $x=\rho/r_0$. By expanding both fractions for small $x$, we finally get the form for the harmonic potential:
\begin{equation}
U(x)=\frac{4}{9}\frac{\beta^3}{\alpha^2}x^2-\frac{4}{27}\frac{\beta^3}{\alpha^2}
\end{equation}
Comparing the above equation with the classical oscillator potential ($m=1$):
\begin{equation}
U(x)=\frac{1}{2}\omega^2x^2+U_0
\end{equation}
and using Eq. (12) we get $\omega^2 \propto P$.

Hence, the pollen tube oscillation frequency at the limit of small oscillations equals $\omega \propto \sqrt{P}$, where $P$ is the turgor pressure (compare also with Fig. 7, where the proportionality  constant ($A$) is estimated from experiment).
\subsection*{Appendix 2}
Pollen tube oscillation: local deflection/wall stress/relaxation/recovery 
\\
\newline
\noindent
STAGES of one oscillation (mechanical view):
\begin{enumerate}
\item Recovery through wall building at $\Gamma$ -- interface: visco-plastic process (elastic equations do not apply here, because of wall and mass production; the system is merely plastic and both subdomains are weakly coupled from the mechanical point of view).
\item Strain and deformation production on $\Gamma$ (the equations apply).
\item Wall stress production on both sides of $\Gamma$ and the resulting strain energy: visco-elastic process (equations apply).
\item Elastic strain energy relaxation in one cycle to produce elongation of one wave length λ (compare Rojas et al, 2011, Fig. 1C, Fig 6A): the phenomenology applies $f =A \sqrt{P}$, finding confirmation in comparison with authors' (Haduch and Pietruszka, 2012) performed experiment, Fig. 7. The process repeats: next oscillation takes place (go to 1. to start another oscillation).
\end{enumerate}

\newpage
\begin{figure}
\begin{center}
\includegraphics[angle=0,scale=.7]{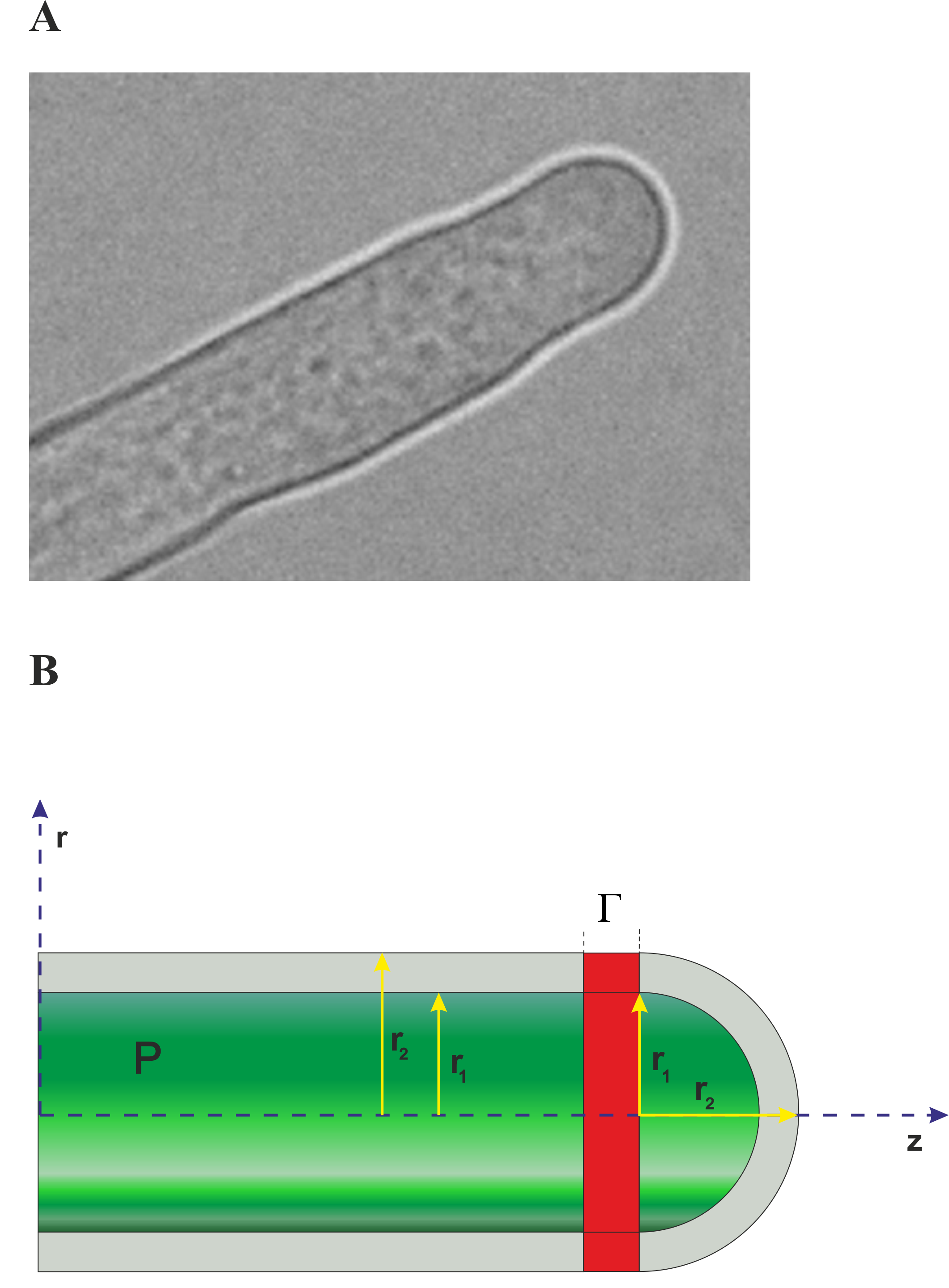}
\caption{\footnotesize {{\em Nicotiana tobaccum} pollen tube apical region. (A) Microscopic view (B) Schematic view: radii of curvature $r_1$ and $r_2$, turgor pressure $P$ and the investigated partition into two distinct regions (a narrow transition zone) are indicated in the chart.
}}
\end{center}
\end{figure}

\begin{figure}
\begin{center}
\includegraphics[angle=0 ,scale=.32]{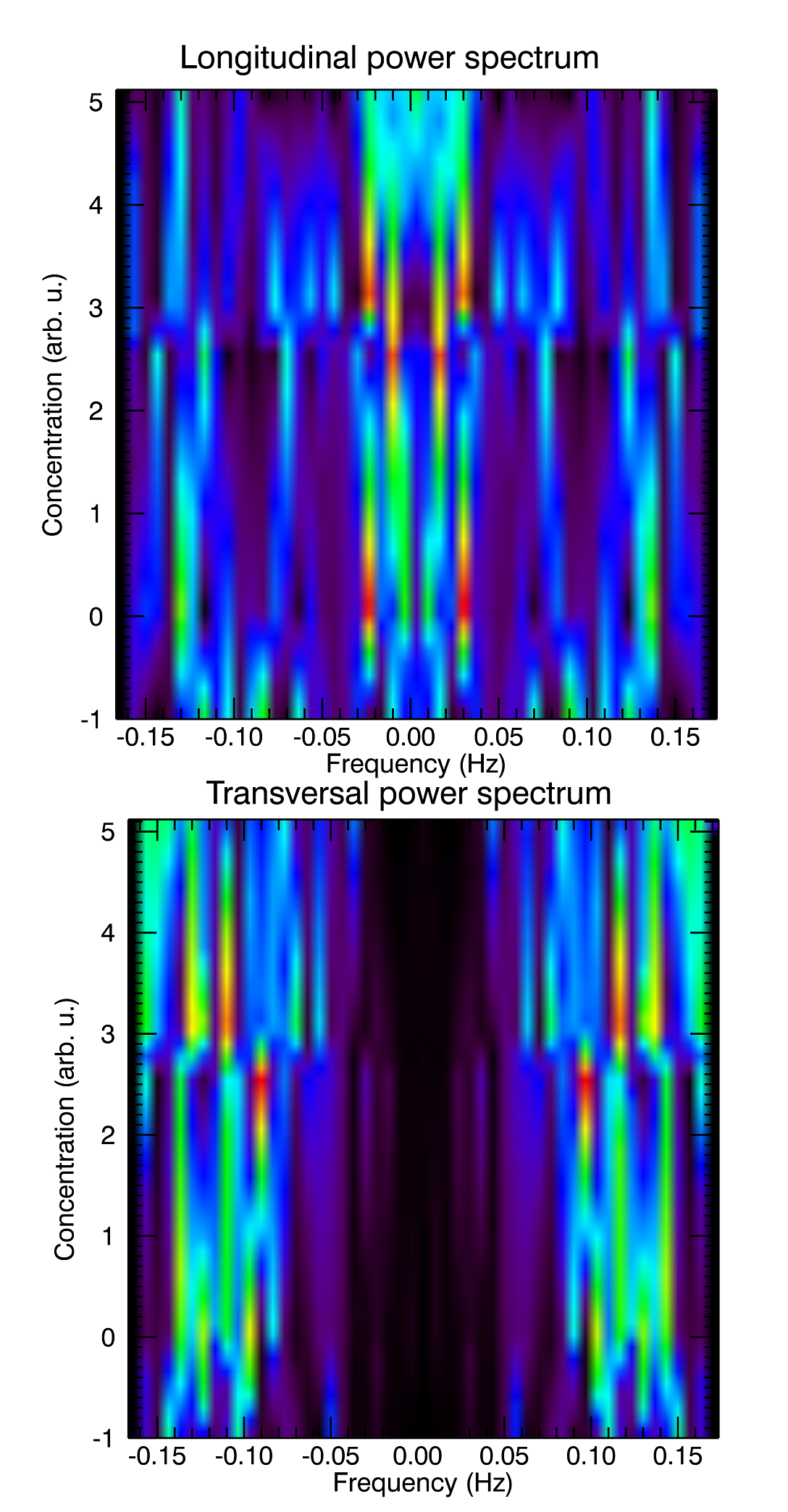}
\caption{\footnotesize {Density plots of the longitudinal and transversal power spectrum of {\em Nicotiana tabacum} pollen tube obtained from the raw experimental data, bias subtracted (Haduch and Pietruszka, 2012) by Fourier analysis,
calculated by the power of the Nyquist criterion and Nyquist rate,
 for different osmotic environments (-1 corresponding to the hypotonic case, 0 -- isotonic case, and 2.5, 3 and 5 corresponding to 25, 30 and 50 mM NaCl in  hypertonic conditions, respectively). 
A broad, narrowing valley at the centre of the lower plot is clearly visible -- low frequencies seen for longitudinal modes are shifted outwards. Red colour indicate high intensity peaks. Interpolated by DAVE, developed at NIST (Azuah et al., 2009).
 }}
\end{center}
\end{figure}

\begin{figure}
\begin{center}
\includegraphics[angle=0,scale=.5]{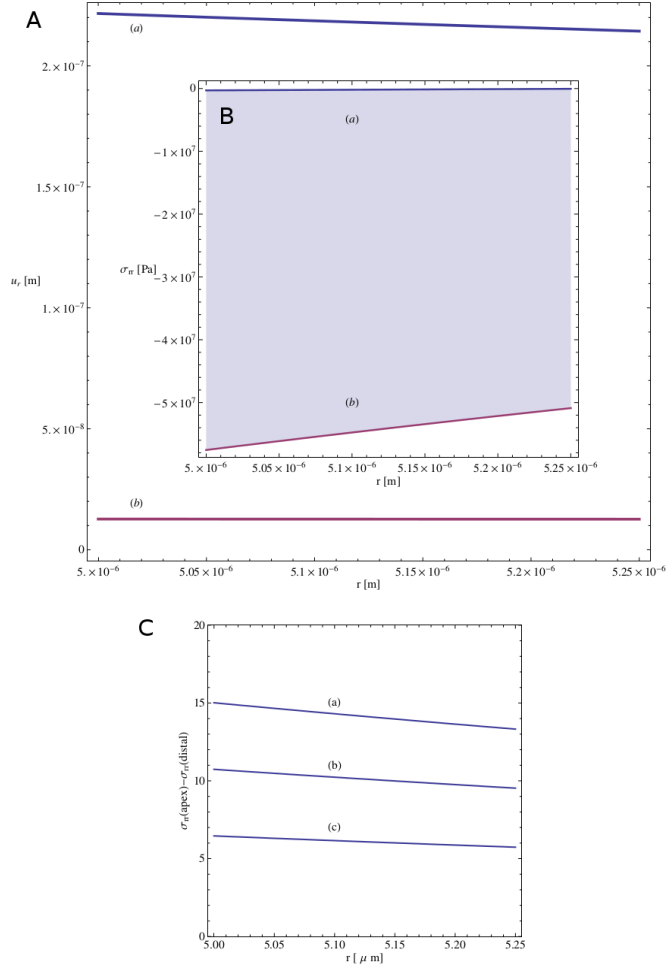}
\caption{\footnotesize {(A) Displacement $u_r$ due to the effective turgor pressure $\tilde{P}$ ($P=0.3$ MPa, $p_{\mathrm{ext}}=0.05$ MPa) acting on the cell wall as a function of the radial distance  $r$ from the pollen tube long axis. Different Young modulus $ \epsilon $ and Poisson coefficient $ \nu $ in both subsystems: (a) 
$u_r=u_r^s$ for a hemispherical apex: $ r_1=5\: \mu$m,
$ r_2 = 5.25\: \mu$m, 
$ \nu = 0.4 $,     
$ \epsilon = 0.2 $ [GPa] (b) $u_r=u_r^c$ for a cylindrical distal part: $ r_1=5\: \mu$m, $ r_2 = 5.25\: \mu $m, 
$ \nu = 0.2$,     
$ \epsilon = 1 $ [GPa]. (B) Tensile stress $\sigma_{rr}$ due to the effective turgor pressure $\tilde{P}$ acting on the cell wall at the position where (a) the cylinder (shank) joins (b) the hemisphere (apex) as a function of the radial distance $r$ from the pollen tube axis. Radial stress discontinuity between the distal wall and apical wall is proportional to the distance between the curves (a) and (b). The calculated maximum wall stress reaches about 60 MPa for the simulation parameters. The strain energy leading to oscillations is proportional to the shaded area. 
(C) Tensile stress difference $\sigma_{rr}(\mathrm{apex})-\sigma_{rr}(\mathrm{distal})$ (in MPa) at the apex  and the distal part parametrized by the changing turgor pressure $P$ acting on the cell wall: (a) $ P=0.5 $ MPa, (b) $ P=0.4 $ MPa and (c) $ P=0.3 $ MPa. 
}}
\end{center}
\end{figure}

\begin{figure}
\begin{center}
\includegraphics[angle=0,scale=.75]{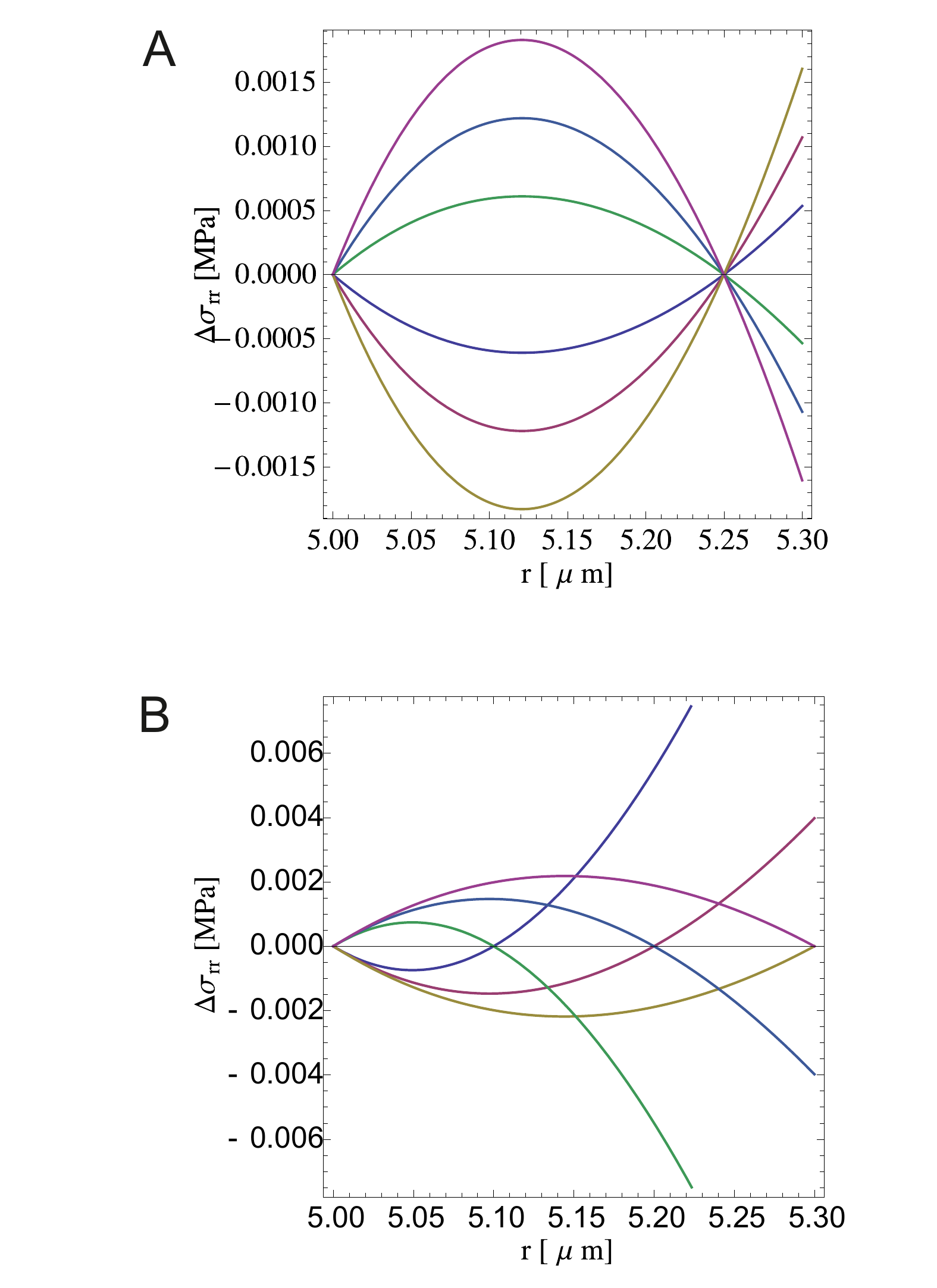}
\caption{\footnotesize 
Tensile stress difference (A) $\sigma_{rr}^c-\sigma_{rr}^s$, upper curves, and the opposite case $\sigma_{rr}^s-\sigma_{rr}^c$, lower curves, calculated at the boundary zone between the approximately hemispherical apical and the cylindrical distal part of the growing pollen tube. Parametrisation by the turgor pressure $P$ acting on the cell wall (upper plots): $ P=0.1 $ (green), $ P=0.2 $ (blue) and $ P=0.3 $ MPa (violet). Remaining parameters for the respective wall geometries: $r_1=5$ $\mu$m, $r_2=5.25$ $\mu$m. 
(B) $\sigma_{rr}^c-\sigma_{rr}^s$, upper curves and the opposite $\sigma_{rr}^s-\sigma_{rr}^c$, lower curves, calculated at the boundary zone between the semispherical apical and the cylindrical distal part of the growing pollen tube, parametrized by the wall thickness (upper plots): $r_2=5.1$ $\mu$m (green), $r_2=5.2$ $\mu$m (blue) and $r_2=5.3$ $\mu$m (violet). The inner wall radius: $r_1=5$ $\mu$m; turgor pressure $ P=0.3 $ MPa.
}
\end{center}
\end{figure}

\begin{figure}
\begin{center}
\includegraphics[angle=0,scale=.6]{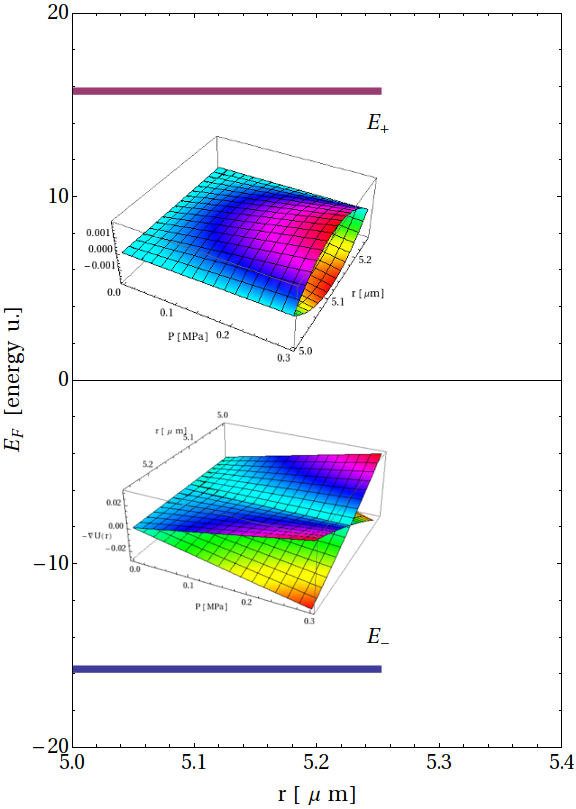}
\caption{\footnotesize 
Frustration energy $E_F$ splitting due to topological effects (Insets: bifurcation diagram and tensegrity force $F= - \nabla U(r)$ balance diagram) in a quasi--2D biological system -- pollen tube apical region. Corresponding symmetry exchange takes place between the resonating residual energy levels $E_{-} \pm \delta E$ and $E_{+} \pm \delta E$ of different major symmetry. Calculation performed at the transition zone between the (hemi-spherical) apical and the (cylindrical) distal part of a growing pollen tube (see Eq. (7)), at a constant turgor pressure $P=0.3$ MPa. The inner and the outer wall radius in both subsystems read: $r_1=5$ $\mu$m, $r_2=5.25$ $\mu$m (wall thickness $\sim 250 $ nm), respectively. The dispersion of each energy level $\delta E\cong 0.0003$.}
\end{center}
\end{figure}

\begin{figure}
\begin{center}
\includegraphics[angle=0,scale=.7]{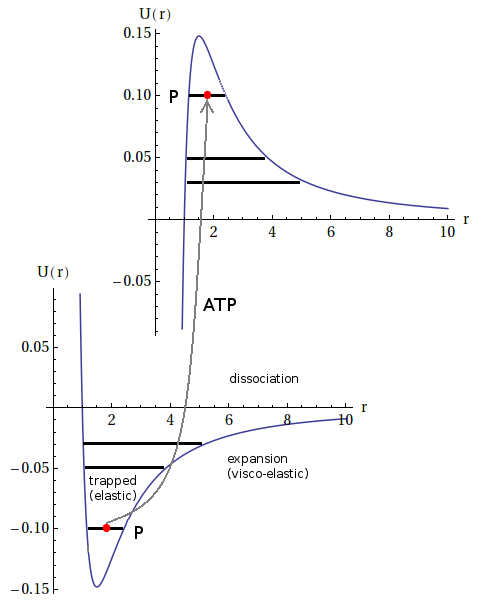}
\caption{\footnotesize {Two branches of the anharmonic potential energy  (frustration potential)  $U_{\pm}(r) \propto \pm \frac{\alpha}{r^3} \mp \frac{\beta}{r^2}$, 
possibly leading to pollen tubes growth rate oscillations, as a function of wall constituing molecules separation $r$. (Here: $\alpha=\beta=1$; in general the coefficients $\alpha$ and $\beta$ are linear in $P$, see Appendix 1).
Oscillations take place between $U_{-}$ and $U_{+}$ potential energy level (by tunneling through symmetry change) yielding  $\omega$, while the amplitude is determined by the actual pressure $P$ level, in accord with experiments (Kroeger and Geitmann, 2011a; Kroeger et al., 2011).
 }}
\end{center}
\end{figure}
\newpage

\begin{figure}
\begin{center}
\includegraphics[angle=0,scale=.525]{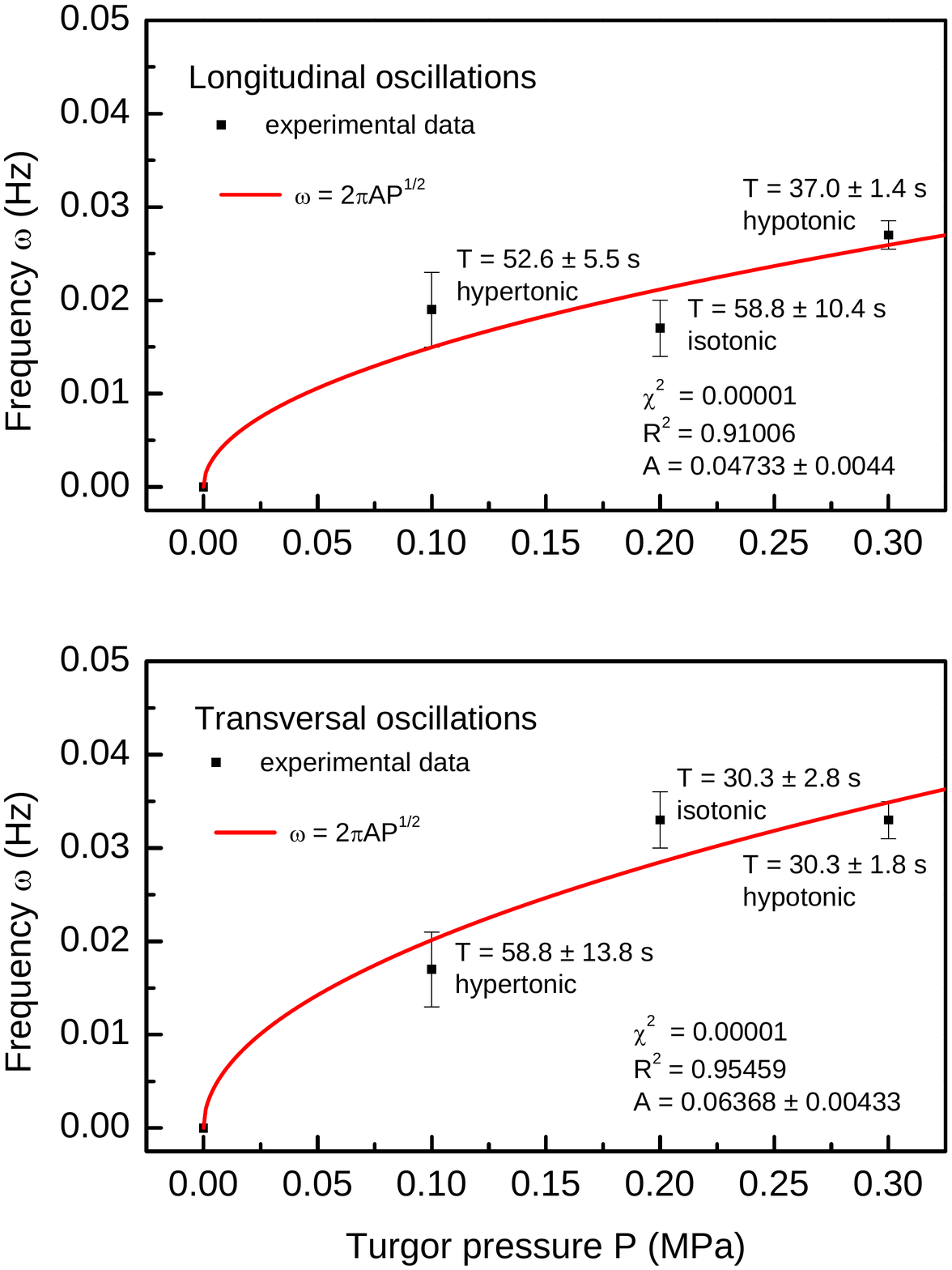}
\caption{\footnotesize { 
Least square fit of the experimental data ($\omega$)
as a function of turgor pressure $P$
 for hypertonic (25 mM NaCl), isotonic and hypotonic (hypo-osmotic stress induced by the addition of water to the gel cultures) treatment of {\em Nicotiana tobaccum} pollen tube (Haduch and Pietruszka, 2012) fitted to the square root function (Appendix 1) derived in this paper ($\omega=2 \pi A \sqrt{P}$; $[A^2]=\mathrm{m/kg}$). 
Stable turgor values correspond to those ranging between 0.1 and 0.4 MPa, which has been recorded using a turgor pressure probe,  (Benkert et al., 1997). The initial point $(0,0)$ added 'by hand' in the chart is exact -- the pollen tube will not oscillate ($\omega=0$) for the vanishing turgor pressure ($P=0$). 
}}
\end{center}
\end{figure}

\end{document}